\documentstyle[prl,aps,epsfig]{revtex}
\begin{document}
\twocolumn[\hsize\textwidth\columnwidth\hsize\csname@twocolumnfalse\endcsname
\title{Longitudinal magnetic excitations in classical spin systems}
\author{Alex Bunker$^*$ and D. P. Landau}
\address{Center for Simulational Physics, University of Georgia, 
Athens, Georgia 30602-2451}
\date{\today}
\maketitle

\begin{abstract}
Using spin dynamics simulations we predict
the splitting of the longitudinal spin wave peak in
all antiferromagnets with single site anisotropy into two peaks
separated by twice the energy gap at the Brillouin zone center. 
This phenomenon has yet to be observed
experimentally but can be easily investigated through neutron scattering 
experiments on MnF$_2$ and FeF$_2$. We have also determined that for
all classical Heisenberg models the longitudinal propagative
excitations are entirely multiple spin-wave in nature.
\end{abstract}

\vskip1pc]
\narrowtext

The mechanism for longitudinal excitations in high
spin magnets with weak to moderate anisotropy,
like MnF$_2$, FeF$_2$, RbMnF$_3$, EuO, and EuS,
is not completely understood. There are conflicting theoretical 
predictions~\cite{Vaks,Mitchell} and
experimental results of limited resolution~\cite{Mitchell,Dietrich,Cox};
however the spin dynamics simulation technique is able to analyze both
the transverse and longitudinal components of the dynamic structure
factor for simple classical Heisenberg models. This is true in both
the hydrodynamic and critical temperature regimes and,
unlike mode coupling theory,
the accuracy of our results can be improved
continuously through the use of more computer time. 
Indeed using high speed supercomputers
we have already achieved considerably higher precision than existing experimental
results.

The above materials all have spin values ($S\ge 2$) which 
are large enough to be effectively described by the classical limit,
$S\rightarrow\infty$,
and bi-linear exchange interactions between nearest, and in some
cases second neighbor atoms on simple lattice structures.
RbMnF$_3$ is an SC antiferromagnet, MnF$_2$ and FeF$_2$
are BCC anisotropic antiferromagnets with weak and moderate
anisotropy respectively, and EuO and EuS are ferromagnets.
The degree of anisotropy in EuO, EuS, and RbMnF$_3$ is negligible. 
While for EuO and EuS, and to a lesser extent MnF$_2$ and FeF$_2$,
the second neighbor interactions are not negligible, a qualitative
understanding of the magnetic dynamics can still be obtained 
through a model with only nearest neighbor interactions. 

We performed our simulation on the isotropic SC Heisenberg magnet
with both ferro- and antiferromagnetic bi-linear interactions
and the anisotropic BCC Heisenberg antiferromagnet.
The Hamiltonian for our model is given by
\begin{equation}
{\cal H} = J\sum_{\langle {\bf rr'}\rangle }{\bf S}_{\bf r} \cdot {\bf S}_{\bf r'}
- D \sum_{\langle {\bf r}\rangle }({\bf S}_{\bf r}^{z})^2
\end{equation}
where  $S_{\bf r}$ is a three-dimensional
classical spin of unit length, $\langle {\bf rr'}\rangle $ 
is a nearest neighbor pair, and
$D$ is the uniaxial single-site anisotropy term.
We determined the dynamic structure factor in the 
[100], [110], and [111] reciprocal lattice directions. 
For the antiferromagnetic case
we have made the transformation $q \rightarrow q + Q$ where
$Q$ is the Brillouin zone boundary in the $[111]$ direction.

We have used the spin dynamics simulation technique, which has been
developed in previous work~\cite{Chen,Bunker}, to calculate
the dynamic structure factor. The spin dynamics simulation technique
involves the creation of an equilibrium distribution of initial
configurations using the Monte Carlo (MC) technique which are then
precessed through constant energy dynamics to yield
the space-time correlation function from each configuration.
These are averaged together and Fourier transformed to obtain a  
result for the dynamic structure factor. By including more
initial configurations the accuracy can be improved indefinitely.
We have used periodic boundary conditions and 
studied lattices of linear sizes of $L = 12$
and $L = 24$. The critical temperatures were determined
using the fourth order cumulant crossing technique~\cite{Binder}
for the anisotropic systems, and $T_c$ is already accurately known for the
isotropic Heisenberg model~\cite{Chen2}. Anisotropy value $D = 0.0591$ was used
for MnF$_2$ to match the experimentally determined~\cite{Als} 
degree of anisotropy.

For all these models we performed the simulation at temperature $T = 0.5T_c$,
low enough to be completely outside the critical regime but not too low
for the MC simulation to produce an equilibrium distribution
of configurations. We also studied $T = 0.8T_c$ and $T = 0.9T_c$ for the
isotropic ferromagnet to investigate the approach to the critical
regime. The results from the higher temperature simulations were
convoluted with a Gaussian resolution function of width
$\delta_\omega$ to minimize effects due to the finite time cutoff.
For the isotropic case we used a distribution of
$1000$ initial configurations for $L = 12$ and $200$ configurations
for $L = 24$, and for the anisotropic case
we used $6000$ configurations for $L = 12$ and $400$ configurations 
for $L = 24$. A smaller number of configurations was
used for $L = 24$ due to limits of computer time.

\begin{figure}
\centerline{\epsfig{figure=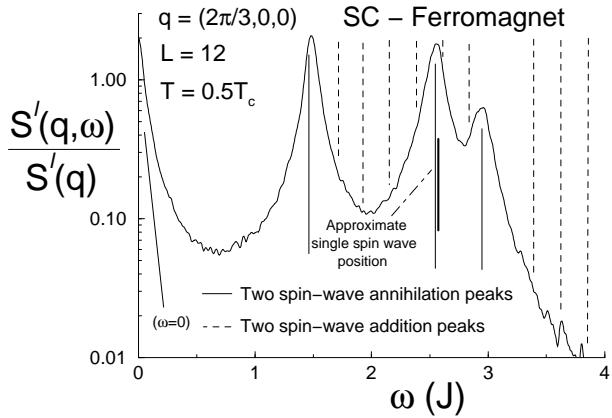,width=8.0cm}}
\vspace*{0.1in}
\caption{The longitudinal component of the
dynamic structure factor for the isotropic
SC ferromagnet vs. frequency. The predicted positions
of the two-spin-wave peaks are superimposed on the
graph of the dynamic structure factor.
Note that a logarithmic scale
has been used for the dynamic structure factor.
}
\label{conf}
\end{figure}

For the longitudinal component of the dynamic structure factor
in  the ferromagnetic case we observed many excitation peaks,
as seen in Fig.~\ref{conf},
however a different set is present for the $L = 12$ and $L = 24$
lattice sizes at the same {\bf q} value. We conjecture that
these excitations are two-spin-wave peaks since
due to finite lattice size effects the frequencies of the
peaks will be limited to certain values
which will be different for different lattice sizes.
In order to test this hypothesis we need to be able to
predict the expected positions of the two-spin-wave peaks.
This requires that we obtain an approximate estimate 
for the general form of the dispersion curve
at the temperature at which the simulation is performed, i. e.
$\omega({\bf q})$ at all points on the reciprocal lattice.

The spin wave stiffness coefficient, $D(T)$, is determined 
from the low $q$, low temperature
limit of the dispersion curve where $\omega = D(T)q^2$ for the
ferromagnetic case and $\omega = D(T)q$ for
the antiferromagnetic case. We assume
that this holds above the low temperature limit, and
the dispersion curve will be given by the linear spin-wave dispersion 
curve, $\omega({\bf q})$ at $T = 0$, 
multiplied by the factor $D(T)/D(0)$. 
At $T = 0.5T_c$ 
the prediction diverges from the real result
only slightly in the high $q$ region where the prediction is at a
lower $\omega$ value than the actual result.
With increasing temperature our approximation of the form
of the dispersion curve becomes less accurate.

Now that we have determined that the linear spin-wave result
multiplied by the factor $D(T)/D(0)$ is a good approximation to
the dispersion curve at $T = 0.5T_c$,
we can estimate the spin-wave frequency over 
all ${\bf q}$ values, not just those
in the reciprocal lattice directions we have measured.
All two-spin-wave creation peaks will thus be at frequency 
\begin{equation}
\omega_{ij}^+({\bf q_i} \pm {\bf q_j}) 
= \omega({\bf q_i}) + \omega({\bf q_j})
\end{equation}
and the spin-wave annihilation peaks will thus be at
frequency 
\begin{equation}
\omega_{ij}^-({\bf q_i} \pm {\bf q_j})
= \omega({\bf q_i}) - \omega({\bf q_j})
\end{equation}
where ${\bf q_i}$ and ${\bf q_j}$ are the wave vectors of the two
spin waves which comprise the two-spin-wave excitation.
As shown in Fig.~\ref{conf}, for the isotropic ferromagnet we see 
a near perfect match of peaks in $S({\bf q},\omega)$ and the predicted
positions of the two spin-wave annihilation peaks, clearly indicating that the
excitations in the longitudinal component are  
due to two-spin-wave annihilation. 

\begin{figure}
\centerline{\epsfig{figure=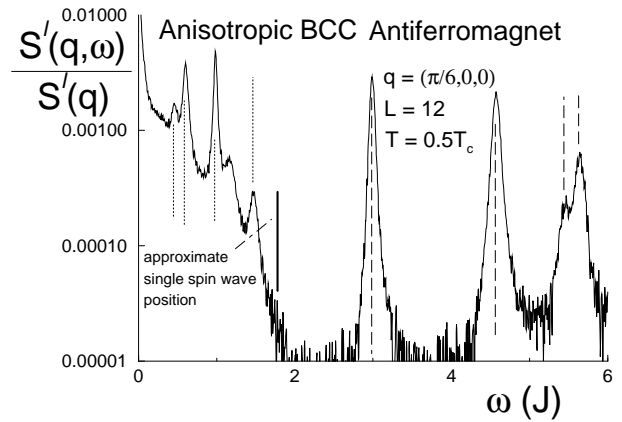,width=8.0cm}}
\vspace*{0.1in}
\caption{The longitudinal component of the
dynamic structure factor for the weakly anisotropic
antiferromagnet, MnF$_2$ vs. frequency.
We compare the predicted positions
of two-spin-wave peaks to the actual peak positions.
The dotted lines are predicted positions
of two-spin-wave annihilation peaks and the dashed lines are
the predicted positions of two-spin-wave creation peaks.
Note, that once again we have used a logarithmic scale.}
\label{lonani}
\end{figure}

For the anisotropic antiferromagnet the approximation
of the dispersion curve from a measurement of $D(T)$
is inaccurate since
the actual dispersion curve does not follow the
functional form of the zero temperature dispersion curve.
Instead we identified a set of two-spin-wave annihilation and
creation peaks which involve spin-waves exclusively
in the reciprocal lattice directions which we measured.
Plotting the expected frequencies of these two-spin-wave
peaks we see a good match between the longitudinal
spin-waves and the expected values for both the annihilation and creation peaks.
This result for weak anisotropy (MnF$_2$) is shown in Fig.~\ref{lonani}.
Note that no trace of a peak is seen at the location of the single
spin-wave peak in the transverse component.
We see, as expected, the presence of both creation and annihilation
spin-wave peaks for the isotropic antiferromagnet as well.

\begin{figure}
\centerline{\epsfig{figure=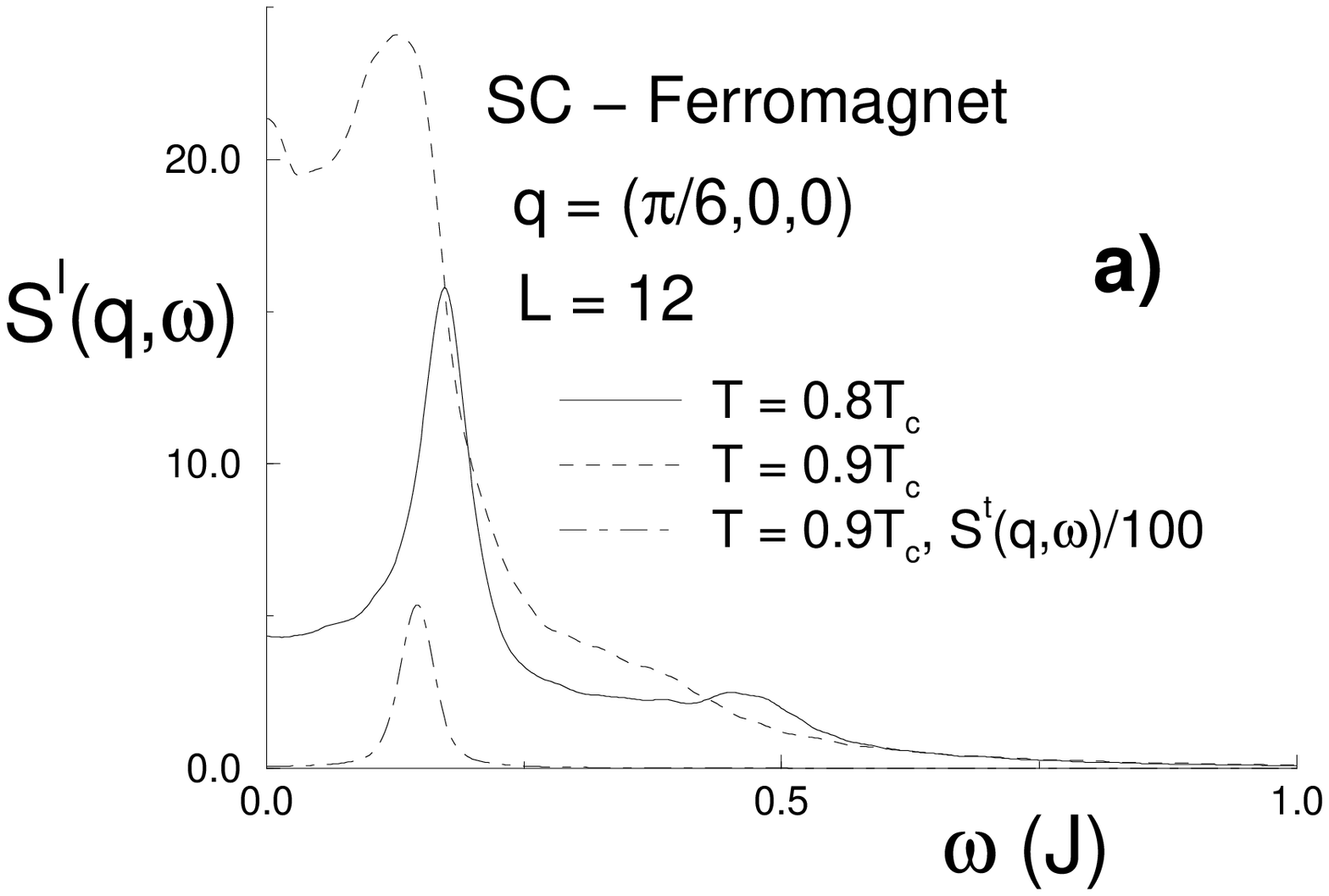,width=8.0cm}}
\centerline{\epsfig{figure=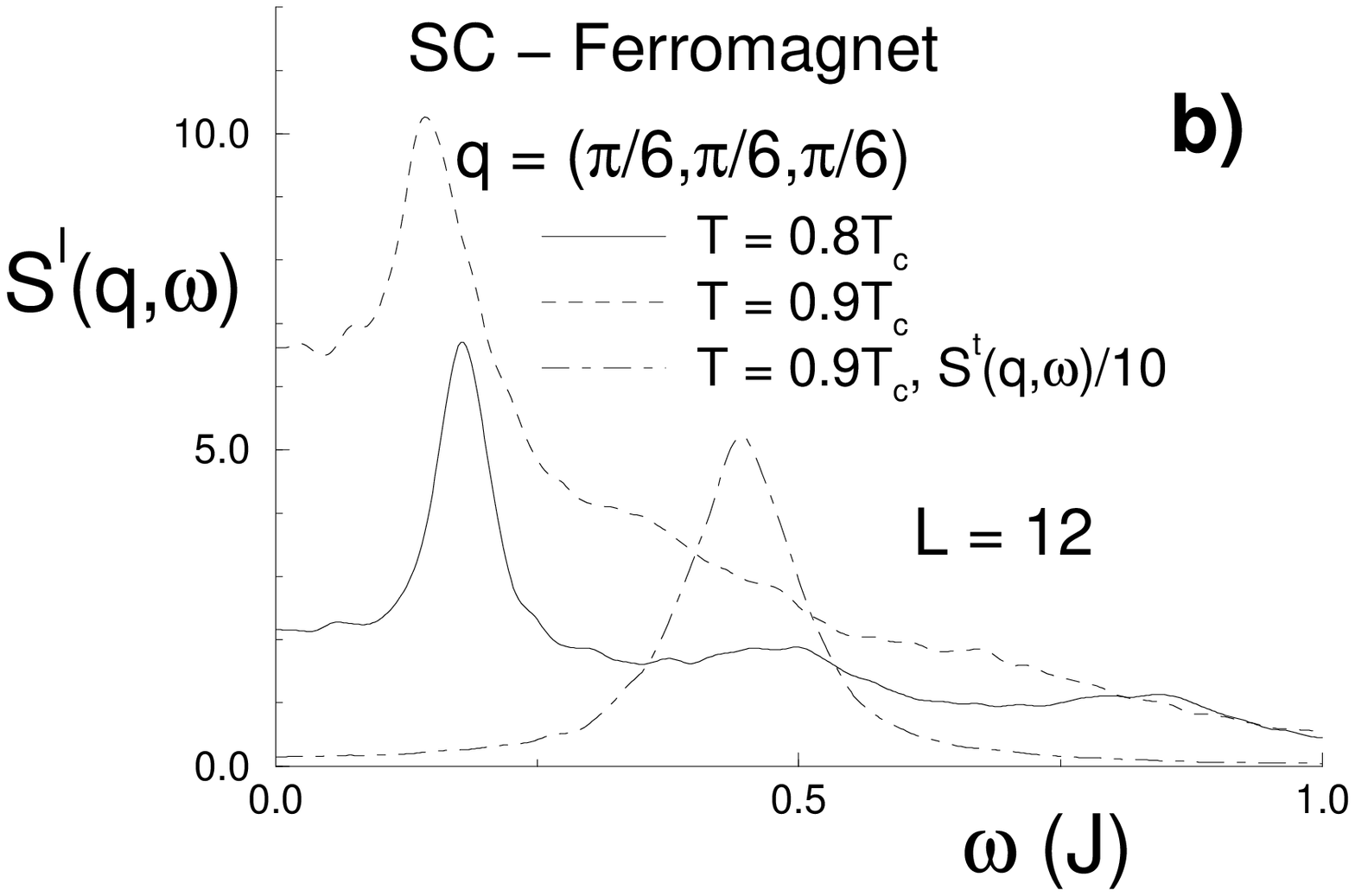,width=8.0cm}}
\vspace*{0.1in}
\caption{The longitudinal and transverse components of the dynamic
structure factor vs. frequency as temperature approaches the 
critical regime are shown; a) shows the [100] direction and b) shows
the [111] direction. A resolution function with coefficient
$\delta_\omega = 0.01/|J|$ was used. The data are unnormalized to show
the relative intensities of the longitudinal and transverse spin-wave
peaks.}
\label{chenwrong}
\end{figure}

The two-spin-wave peaks where one of the single spin-waves
of which they are made up has the 
lowest ${\bf q}$ are the most intense.
This is to be expected since the lower ${\bf q}$ single spin-waves
are more intense themselves.
As the temperature rises the two-spin-wave peaks broaden.
For the antiferromagnetic case, where the longitudinal
and transverse components can not be separated, as $T$ approaches $T_c$
the two-spin-wave peaks disappear into the
tails of the single spin-wave peak and the diffusive
central peak. For the ferromagnetic case the two-spin-wave peaks
do not disappear entirely, the two-spin-wave peak
corresponding to the lowest ${\bf q}$ spin-waves
remains and the other two-spin-wave peaks
broaden to disappear into its tail.
A previous study by Chen et al.\cite{Chen}
misidentified the peak in the longitudinal component
as a single spin-wave excitation.
This happened because they were only able to look
in the [100] lattice direction for which the dominant two-spin-wave
is at the same frequency as the single spin-wave peak
for any given ${\bf q}$. Since for ferromagnets
only spin-wave annihilation peaks are present, the 
dominant two-spin-wave process found in the [100] direction is [q,0,1]-[0,0,-1].
Since in the low ${\bf q}$ limit $\omega \propto q^2$,
$\omega([q,0,0]) \approx \omega([q,0,1]) - \omega([0,0,-1])$.
This is however not the case for the [111] direction.
In Fig.~\ref{chenwrong} we show the longitudinal and transverse components
in the [100] and [111] directions at temperatures 
approaching $T_c$. Note the fact that the longitudinal
and transverse spin wave peaks are at the same frequency
in the [100] direction but not in the [111] direction. 

The existence of a set of finite two-spin-wave peaks
is an artifact of the finite
size of the lattice we use in our simulation.
If one were to measure the longitudinal component
of the dynamic structure factor for a real crystal, i. e. 
effectively an infinite lattice of magnetic moments,
the spectrum of possible two-spin-wave peaks would be
continuous.
The longitudinal component of the dynamic structure factor
for the isotropic antiferromagnet has been measured experimentally
by Cox et al.~\cite{Cox}. Taking measurements in the [111] direction,
they found a diffusive central peak
and a propagative spin-wave peak at the same frequency as the
transverse single spin-wave peak. The longitudinal component
of the dynamic structure factor for the isotropic ferromagnet
EuO has been found experimentally to have a propagative
spin-wave and no diffusive central peak~\cite{Dietrich}.

For both the ferro- and antiferromagnetic cases,
the longitudinal excitations at the single
spin-wave peak frequency can be explained in terms of 
two-spin-wave peaks as follows. Since spin-wave intensity
increases with decreasing $q$ the most intense two-spin-waves 
will be those which are comprised of one spin-wave with a very
small $q$ and another spin-wave with a ${\bf q}$ very close to the
resultant two-spin-wave ${\bf q}$. If we look at the limit of infinite
lattice size, i.e. the real system,
for the isotropic case a spin-wave with extremely small $q$ 
will have a negligible frequency.
Thus the intensity of the two-spin-wave creation and annihilation
peaks will have a maximum at the single spin-wave frequency.
Even though only annihilation two-spin-wave peaks are present
for the ferromagnetic case one should still see this effect
in both the ferro- and antiferromagnetic cases and this is
what is seen in the experiments~\cite{Dietrich,Cox}. 
We see no evidence of a diffusive
central spin-wave peak in the longitudinal 
component of the dynamic structure factor for the isotropic ferromagnet 
in agreement with the experimental results of Dietrich et al.~\cite{Dietrich}, 
and the theoretical result of Villain~\cite{Villain} 
but in disagreement with the theoretical predictions of
Vaks et al.~\cite{Vaks}.

When we apply this same reasoning to the anisotropic antiferromagnet,
which as shown in Fig.~\ref{lonani} also displays the same 
two-spin-wave peak behavior,
we are left with an extremely intriguing result which can
be measured experimentally. A spin-wave at very small ${\bf q}$ will
no longer have a negligible frequency but instead the frequency
of the energy gap at the Brillouin zone center. If our hypothesis 
about the origin of excitations in 
the longitudinal component of the dynamic structure factor is correct then
for an anisotropic antiferromagnet 
one will observe an apparent splitting in the
spin-wave peak in the longitudinal component. 
In an infinite lattice there will be a peak due to two-spin-wave
creation which is shifted upwards from the single spin-wave
frequency by an amount equal to the energy gap,
and a spin-wave annihilation peak shifted downwards by the same frequency.

The theoretical explanation as to why the longitudinal 
component of the dynamic structure factor is made up of 
two-spin-wave peaks and why only two-spin-wave annihilation
peaks are present for the ferromagnetic case while
both two-spin-wave annihilation and creation peaks
are present for the antiferromagnetic case is as follows.
If we express the dynamics in quantum mechanical
formalism the spins, as they appear in the Hamiltonian, 
are expressed in terms of the operators ${\bf S}_i^+$, 
${\bf S}_i^-$, and ${\bf S}_i^z$ where
\begin{eqnarray}
{\bf S}_i^+ &=& {\bf S}_i^x + i{\bf S}_i^y \\
{\bf S}_i^- &=& {\bf S}_i^x - i{\bf S}_i^y. 
\end{eqnarray}
${\bf S}_i^+$ and ${\bf S}_i^-$ can be expressed
in terms of ladder operators, 
$a_i$ and $a_i^+$, which
raise or lower ${\bf S}_i^z$
by one quanta.
In the linear approximation
\begin{eqnarray}
{\bf S}_i^+ &=& (2{\bf S})^{1/2}a_i^{\dag}\\
{\bf S}_i^- &=& (2{\bf S})^{1/2}a_i.
\end{eqnarray}
If we take the approximation for ${\bf S}_i^z$ to one order
higher than ${\bf S}_i^z = {\bf S}$ then
\begin{equation}
{\bf S}_i^z = {\bf S} - a_i^{\dag}a_i = {\bf S} -
\frac{1}{2{\bf S}}{\bf S}_i^+{\bf S}_i^-.
\end{equation}
For the classical limit this corresponds to an
infinitesimal raising and lowering of the
$z$ component of the spin which is a longitudinal 
spin-wave. These spin-waves will be at the frequencies
corresponding to the difference between the frequencies
of single transverse spin-waves since they result from
creation followed by annihilation of a spin-wave.
Thus for the ferromagnetic case all two-spin wave excitations
will result from these annihilation processes.

Unlike the ferromagnetic the antiferromagnet is not a quantum 
state of the Heisenberg Hamiltonian.
As a result the $a_i$ and $a_i^+$ are
replaced through a Bogoliubov transformation
by new operators which are a linear combination of the
creation and annihilation operators but which only connect
excitations on the same sublattice~\cite{Mattis}. As a result both
creation and annihilation two-spin-wave excitations exist.

In conclusion, for magnets
which can be described by a classical spin
model, the longitudinal propagative
excitations are made up of two-spin-wave peaks. 
In agreement with our theoretical picture only annihilation 
two-spin-wave peaks were present for the ferromagnetic case
but both annihilation and creation two-spin-wave
peaks are present for both the isotropic and
anisotropic antiferromagnets.
As the temperature approached the
critical regime the longitudinal two-spin-wave peaks broadened
and converged into a single peak at the frequency
of the dominant lowest-$q$ two-spin-wave peak.
In an isotropic lattice of infinite size, i.e. a real crystal,
the two-spin-waves will result in a peak at the
single spin-wave frequency for either ferromagnetic or 
antiferromagnetic isotropic systems, 
in agreement with experimental results~\cite{Dietrich,Cox}. 
For the anisotropic antiferromagnet
both creation and annihilation two-spin-wave peaks
exist, and
the longitudinal component of the anisotropic antiferromagnet
should show two peaks; a two-spin-wave annihilation peak
at the single spin-wave peak frequency minus the energy gap 
frequency, and a two-spin-wave creation peak at the
single spin-wave peak frequency plus the energy gap 
frequency. This result indicates the presence of a new form
of excitation behavior in magnetic materials which can be 
directly tested experimentally. 

We thank Shan-Ho Tsai, Roderich Moessner, Werner Schweika,
and Michael Krech For Helpful suggestions and stimulating discussions. 
This research was supported in part by NSF grant \#DMR9727714.

\end{document}